\documentclass[10pt,journal]{IEEEtran}
\usepackage{cite}

\ifCLASSINFOpdf
\else
\fi
\usepackage{graphicx,times,amsmath} 

\usepackage{amsthm}
\usepackage{amssymb}
\usepackage{algorithm} 
\usepackage{multirow}
\usepackage{xcolor}
\usepackage{mathrsfs}
\usepackage{hyperref}
\usepackage{url}

\usepackage{subfigure}
\usepackage{caption}
\usepackage{float}
\usepackage{algorithmic}

\usepackage{balance}
\usepackage{epstopdf}
\usepackage{doi}

\theoremstyle{definition}

\theoremstyle{plain} \newtheorem{theo}{Theorem}   \newtheorem{lemm}{Lemma}

\begin{document}


%
%
%
%
%
%
%
%
%
%
%


%
\title{A Hybrid Optimization Approach to Demand Response Management for the Smart Grid}
%
%
%

\author{Fan-Lin~Meng,~\IEEEmembership{Student Member,~IEEE,}
        and~Xiao-Jun~Zeng,~\IEEEmembership{Member,~IEEE}
\thanks{Fan-Lin Meng and Xiao-Jun Zeng are with the School of Computer Science, University of Manchester, Manchester, United Kingdom (email: mengf@cs.man.ac.uk, x.zeng@manchester.ac.uk).}}

\maketitle

\begin{abstract}


This paper proposes a hybrid approach to optimal day-ahead pricing for demand response management. At the customer-side, compared with the existing work, a detailed, comprehensive and complete energy management system, which includes all possible types of appliances, all possible applications, and an effective waiting time cost model is proposed to manage the energy usages in households (lower level problem). At the retailer-side, the best retail prices are determined to maximize the retailer's profit (upper level problem). The interactions between the electricity retailer and its customers can be cast as a bilevel optimization problem. To overcome the weakness and infeasibility of conventional Karush--Kuhn--Tucker (KKT) approach for this particular type of bilevel problem, a hybrid pricing optimization approach, which adopts the multi-population genetic algorithms for the upper level problem and distributed individual optimization algorithms for the lower level problem, is proposed. Numerical results show the applicability and effectiveness of the proposed approach and its  benefit to  the retailer and its customers by improving the retailer's profit and reducing the customers' bills.
\end{abstract}

\begin{IEEEkeywords}
Smart grid, demand response management, day-ahead pricing, bilevel optimization, multi-population genetic algorithms
\end{IEEEkeywords}

%
\IEEEpeerreviewmaketitle

\section{Introduction}
%
%
%
%

With the large-scale deployment of smart meters and two-way communication infrastructures, the benefits of demand response will be greatly enhanced. Among various demand response programs, time-differentiated pricing models are regarded as promising strategies to balance the load and supply, reduce the peak demands and increase the grid reliability \cite{palensky2011demand}. One typical type of pricing among the time-differentiated pricing models is day-ahead pricing \cite{siano2014demand}, in which customers receive the hourly prices for the next 24 hours.



The existing research on demand response and time-differentiated pricing can be categorized into the following three directions. Firstly, \cite{Leon-garcia2010,Mohsenian-Rad2010,adika2013autonomous,liu2014peak} deal with how customers respond to the time-differentiated prices. Secondly, the work of \cite{Samadi2010,Li2011,yang2012game} are concerned with how the retailers set the electricity prices, where they model the customers' energy consumption preferences in the form of  utility functions. Thirdly, \cite{Yu2011,chen2011innovative,chen2012optimal,qian2013demand,maharjan2013dependable,zugno2013bilevel,chai2014demand}
deal with how the retailers determine the electricity prices based on the expected responses of customers where they model the interactions between a retailer and its customers as a Stackelberg game or bilevel problem. All of the above work and more unreferenced work have provided valuable findings in the demand response management area. Despite these contributions, there are still notable gaps or weaknesses in the existing approaches:

From the customer modelling point of view, firstly the existing research has failed to model certain important type of appliances commonly used in most households. For example, all works given in \cite{Leon-garcia2010, Mohsenian-Rad2010,adika2013autonomous,liu2014peak} have not addressed the usage optimization modelling problem for curtailable appliances (such as air conditioning);
secondly for interruptible appliances and non-interruptible appliances, the existing waiting cost model \cite{Leon-garcia2010} \cite{chen2011innovative} is a pure theoretic one, which is impossible for most ordinary customers to set up and use and therefore inapplicable; thirdly theoretic household utility functions rather than utility functions based on real home appliances are often used like in \cite{Samadi2010,Li2011,yang2012game}. Although these theoretic models provide valuable insight for demand response management and pricing, this type of abstract model cannot be used by customers to find the best usage and scheduling scheme to minimize their bills or maximize their benefits. Noticing these shortcomings, the first motivation of our research is to propose a home appliances based utility function model that includes all possible applications and to develop an applicable and implementable optimal solution usable by ordinary customers to find the best usage and scheduling scheme. 

From the retailer modelling point of view, the pricing optimization problems and models by the existing research are either oversimplified or unrealistic from an application point of view. For example, the pricing optimization problems and models 
only model one or two types of appliances in the customers' level problem \cite{chen2011innovative}\cite{qian2013demand}\cite{zugno2013bilevel} or  fail to give the realistic and explicit form of customers' utility functions \cite{Yu2011}\cite{chen2012optimal}\cite{maharjan2013dependable} \cite{chai2014demand}. In other words, such problem formulations and models are only partially or unrealistically modelling of a retailer's pricing optimization problem and therefore are insufficient. For this reason, the second motivation of our research is to develop a complete and comprehensive pricing optimization model that accurately and realistically represents the real pricing problem faced by retailers to enable the usability and applicability of the resulting models. 

From the day-ahead pricing computation point of view, finding the best day-ahead pricing scheme often requires solving a bi-level optimization problem.
In  current bi-level optimization theory, the common solution is to cast bi-level optimization problems into equivalent single-level optimization problems by replacing the lower level problems with their KKT optimality conditions.
Such a KKT approach has been used by the existing research such as \cite{zugno2013bilevel} and \cite{momberretail} to solve the optimization problems faced by retailers. 
Unfortunately, this approach is infeasible in the large-scale applications when a retailer has thousands to millions of customers, where each customer may have several constraints at the lower-level optimization problems. When using KKT approach, it will result in far too many constraints for the resulting single-level problem which is infeasible to be solved by existing optimization software. 
Another important issue worth mentioning is the customers' privacy concerns. By replacing the lower level problem with its KKT conditions, the lower-level problem will be exposed to the retailer and may cause privacy problems for customers. For the above reasons, the third motivation of our research is to develop a hybrid optimization approach to solve the proposed bilevel problem in a distributed manner to overcome the above weaknesses.


By recognising the weaknesses in the existing methods and following the above three motivations, in this paper, we propose a hybrid approach to maximize customers' benefits and optimize retailers' day-ahead prices for demand response management within the smart grid.

The rest of this paper is organized as follows. The system model is proposed in Section \ref{model} and the bilevel problem formulation is presented in Section \ref{problem formulation}. In Section \ref{model_solution}, the optimal bilevel solution is given. Numerical results are provided and discussed in Section \ref{results}. The paper is concluded in Section \ref{conclusion}.

\section{System Modelling} \label{model} 
In this section, how the optimal day-ahead pricing for demand response management can be modelled as a bilevel optimization problem is described. 

It is assumed that each customer is equipped with a smart meter. The interactions between the retailer and its customers are enabled through a two way communication infrastructure. 

%

The decision processes of the retailer and its customers are: the retailer acting as the upper level decision agent firstly announces the selling price to its customers with the aim to maximize its profit. To solve this profit maximization problem, it is assumed that each customer (lower level decision agent) optimally reacts to the announced retail price, i.e. each customer (smart meter) determines the optimal energy consumption with the aim to maximize its benefits such as minimizing its bills.


Note that each customer's decision about electricity usage is independent from other customers' decisions. As a result, our considered optimization problem faced by the retailer can be seen as a bilevel optimization problem with independent customers.

  
%

The general formulation of a bi-level optimization problem with one upper level decision agent and $N$ independent lower level decision agents can be represented as follows: 
\begin{equation}  \label{bilevel_formualtion}
\begin{array}{ll} 
\max \limits_{x,y_{1},...,y_N}& F(x,y_{1},...,y_N) \\ 
\hbox{subject to} & \\ 
& y_i \in \underset{y_i}{\operatorname{argmin}} \lbrace \hspace{1mm} f_i(x,y_i):g_i(x,y_i)\leq 0, \\
& \quad \quad \quad \quad \quad \quad i=1,\dots,N \hspace{1mm} \rbrace\\ & G(x,y_{1},...,y_N)\leq 0 \\
 & x\in X, y_i\in Y_i 
\end{array}
\end{equation}


In the above formulation, $F$ represents the upper-level objective function and $f_i \hspace{0.1cm}   (i=1,2,...,N)$ represent the lower-level objective functions. Similarly, $x$ is the decision vector of the upper level agent and $y_i$ is the decision vector of the $i$-th lower level agent. $G$ represents the constraint functions at the upper level and $g_i$ represents the constraint functions of the $i$-th lower level agent.  $X$ are the bound constraints for the upper level decision vector and $Y_i$ are the bound constraints for the decision vector of the $i$-th lower level agent.

A solution $(x^*, y_1^*,...,y_N^*)$ which maximizes the above objective function $F(x,y_{1},...,y_N)$ subject to all the constraints is said to be a bilevel optimal solution.



%

\section{Bilevel Problem Formulation} \label{problem formulation}

In this section, the mathematical representation of the considered two level decision making problems is provided. Firstly, our focus is to formulate the energy management problem in response to the day-ahead pricing in each household at the lower level. Secondly, we model the profit maximization problem for the retailer at the upper level who will offer the 24 hours prices to its customers. 

Throughout this paper, let $ \mathcal{N} \triangleq \{1,2,...,N\}$ denote the considered set of customers with $N \triangleq | \mathcal{N} |$, and  $\mathcal{H} \triangleq \{1,2,...,H\}$ where $H = 24$ denotes the scheduling horizon. We define the prices offered by the retailer as a price vector: $P=[p^1,...,p^h,...,p^H]$, where $p^h$ represents the electricity price at hour $h$.

\subsection{Customer-side Problem at the Lower Level}

%

%



%

We categorize the home appliances into non-shiftable appliances (e.g. lights), interruptible appliances (e.g.  PHEVs), non-interruptible appliances (e.g. washing machines, dish washers) and curtailable appliances (e.g.  air conditioning, heating) according to their load types \cite{meng2014optimal}.

%
%
%

In the following, we will firstly give the mathematical models for the different types of appliances. Furthermore, we propose a financial-incentive based waiting cost model for interruptible and non-interruptible appliances.

For each customer $ n \in \mathcal{N} $ , we denote the set of all appliances in the household as $ A_n $, non-shiftable appliances as $NS_n$, interruptible appliances as $I_n$, non-interruptible appliances as $NI_n$ and curtailable appliances as $C_n$.

\subsubsection{Interruptible Appliances}


For each interruptible appliance $ a \in I_n $, a scheduling vector of energy consumption over the scheduling window $\mathcal{H}= \{1,2,...,H\}$  is defined as $\mathrm{x}_{n,a}=[x_{n,a}^1,...,x_{n,a}^h,...,x_{n,a}^H]$
where $ x^h_{n,a} \geq 0$ represents the $n$-th customer's electricity consumption of appliance $a$ at time $h$. Furthermore, the scheduling window for each appliance $a$ can be set by each customer according to his/her preference and is defined as $\mathcal{H}_{n,a} \triangleq \{\alpha_{n,a}, \alpha_{n,a}+1,...,\beta_{n,a}\}$ \cite{Leon-garcia2010}.  Since the window $\mathcal{H}_{n,a}$ is consecutive, one only needs to specify the beginning
scheduling time $\alpha_{n,a} \in \mathcal{H}$ and end time $\beta_{n,a} \in \mathcal{H}$.


The model of the payment minimization problem for each interruptible appliance is given as follows:
\begin{align}
  &\min J_{I_n(a)}(\alpha_{n,a} \colon \beta_{n,a}) = \min_{x_{n,a}^h} \sum_{h=\alpha_{n,a}}^{\beta_{n,a}} p^h \times x_{n,a}^h \label{1}\\
  &s.t. \nonumber\\
  &\sum_{h=\alpha_{n,a}}^{\beta_{n,a}} x_{n,a}^h=E_{n,a}, \label{2}\\
  &\gamma_{n,a}^{min}\leq x_{n,a}^h \leq \gamma_{n,a}^{max}, \forall h \in
\mathcal{H}_{n,a} \label{3}.
  \end{align}
  
Constraint (\ref{2}) represents that, for each appliance $a$, the total energy consumption to accomplish the operations within the scheduling window  is fixed, which is denoted as $E_{n,a}$. Constraint (\ref{3}) represents that there is a  minimum power level and a maximum power level for each appliance $a$ within the scheduling window. 
%
%

 
\subsubsection{Non-interruptible Appliances} 
As the operations of each non-interruptible appliance $a \in NI_n$ are consecutive, we define the length of the operations $L_{n,a}$. The customers can set the scheduling 
window $ \mathcal{H}_{n,a} \triangleq \{\alpha_{n,a}, \alpha_{n,a} + 1,...,\beta_{n,a}\}$ by specifying the beginning scheduling time and end time.

The optimization problem is to find each appliance's optimal start time $s_{n,a}^*$  to minimize the customer's payment. 

As a result, the model of the payment minimization problem for a non-interruptible appliance is given as follows:
\begin{align}
&\min J_{NI_n(a)}(\alpha_{n,a} \colon \beta_{n,a}) =  \underset{s_{n,a}} {\operatorname{\min}} \{ \underset{x_{n,a}^h} {\operatorname{\min}} \sum_{h=s_{n,a}}^{s_{n,a}+ L_{n,a}}  p^h \times  x_{n,a}^h \} \label{5}\\
 &s.t. \nonumber\\
  &\sum_{h=s_{n,a}}^{s_{n,a}+ L_{n,a}} x_{n,a}^h=E_{n,a}, \label{6}\\
  &\alpha_{n,a} \leq s_{n,a} \leq \beta_{n,a} - L_{n,a}, \label{7}\\
  & \gamma_{n,a}^{min}\leq x_{n,a}^h \leq \gamma_{n,a}^{max},      \forall h \in \mathcal{H}_{n,a}. \label{8}
  \end{align}
  
Constraint (\ref{6}) represents that the total energy consumption to accomplish the consecutive operations is fixed at $E_{n,a}$. Constraint (\ref{7}) indicates that the start time is bounded within the interval $[\alpha_{n,a}, \beta_{n,a}-L_{n,a}] $. Constraint (\ref{8}) shows that there is a minimum power level and a maximum power level for each appliance $a$ within the scheduling window. 

%

\subsubsection{Waiting Time Cost Model}

%
%

We propose a financial-incentive based waiting cost model that is straightforward and easy to use for interruptible and non-interruptible appliances. In real applications, firstly the customers need to set the financial thresholds that trigger different waiting lengths. For example, customers can input their financial thresholds via the interface between a laptop or mobile phone and the home energy management software integrated in a smart meter. Additionally such home energy management software can provide guidance and support in helping customers determine their financial thresholds by providing a questionnaire to customers. With the relevant information available, the customers can set the financial thresholds themselves easily. Secondly, by using our proposed waiting cost model, it will determine the optimal waiting length for each appliance. Example 1 is used to help describe the proposed waiting time scheme.

\textbf{\textit{ Example 1.}} We assume that the original scheduling window for PHEV is [7PM - 11PM] and the maximum waiting time length is 3 hours. Furthermore, the financial thresholds to trigger the waiting are 10 pence for 1 hour, 25 pence for 2 hours and 45 pence for 3 hours. Assume that the energy bills saved by different waiting hours are given in Table \ref{table_threshold}.

\begin{table}[htp] 
\centering
\caption{Energy bills saved by different waiting length}
\label{table_threshold}
\begin{tabular}{|l|l|l|l|}
\hline
Waiting Length & Scheduling Window & {\begin{tabular}[c]{@{}l@{}}Financial \\ Threshold\end{tabular}} & Saved Bill  \\ \hline
0 hour & {[}7PM - 11PM{]} & - & - \\ \hline
1 hour & {[}7PM - 12AM{]} & 10 pence & 12 pence \\ \hline
2 hours & {[}7PM - 1AM{]} & 25 pence & 30 pence \\ \hline
3 hours & {[}7PM - 2AM{]} & 45 pence & 40 pence \\ \hline
\end{tabular}
\end{table}
From the Table \ref{table_threshold}, we can see that by waiting for 1 hour, it can save the  bill by 12 pence, which is higher than the financial threshold (10 pence). As a result, the waiting time cost model will treat the current waiting length (1 hour) as a potential solution and then check the next waiting length (2 hours). Due to the same reason as before, the waiting length of 2 hours can also be treated as a potential solution. However, by waiting for 3 hours, it can only save the customer by 40 pence on the bill that is lower than the financial threshold (45 pence). As a result, the waiting length of 3 hours cannot be regarded as a potential solution. The above process is repeated until all the waiting lengths are checked. The optimal waiting time is one potential solution that achieves the maximal bill saving for the customer. In the above example, the optimal waiting time for the PHEV is 2 hours. 

Based on the above analysis and illustration, the mathematical representations of the waiting time cost model are given below. To avoid repetition, we only deal with the waiting time cost model for interruptible appliances. However, the model also applies to non-interruptible appliances. 

For each interruptible appliance $a \in I_n$, we assume that the maximum waiting time is denoted as $K_{n,a} \geq 1$, which can be set by customers according to their preferences in advance.

Without any waiting time, the minimized energy bill model for each interruptible appliance is denoted as follows:
\begin{equation}
\min J_{I_n(a)}(\alpha_{n,a} \colon \beta_{n,a}) = \min_{x_{n,a}^h} \sum_{h=\alpha_{n,a}}^{\beta_{n,a}} p^h \times x_{n,a}^h
\end{equation}

With the waiting time of  $k_{n,a} \in \{0,...,K_{n,a}\}$, the minimized energy bill model is defined as:
\begin{equation}
\min J_{I_n(a)}(\alpha_{n,a} \colon \beta_{n,a}+ k_{n,a})= \min_{x_{n,a}^h} \sum_{h=\alpha_{n,a}}^{\beta_{n,a} + k_{n,a}} p^h \times x_{n,a}^h, 
\end{equation} 
where $\mathcal{H}_{n,a} =\{\alpha_{n,a}, \alpha_{n,a}+1,...,\beta_{n,a}\}  $ is extended to $ \{\alpha_{n,a}, \alpha_{n,a}+1,...,\beta_{n,a}+ k_{n,a}\} $.

We define the \textit{Waiting Time Benefit Function}, i.e., the energy bill saved by waiting $k_{n,a}$ hours, as follows:
\begin{equation}
\begin{array}{lr}
\triangle J_{I_n(a)}(k_{n,a}) = \min J_{I_n(a)}(\alpha_{n,a} \colon \beta_{n,a}) - \\
\min J_{I_n(a)}(\alpha_{n,a} \colon \beta_{n,a}+ k_{n,a}), k_{n,a} = 0,1,...,K_{n,a}
\end{array}
\end{equation}

Furthermore, we define the \textit{Benefit Threshold Function} as follows: 
\begin{equation}
\triangle \hat{J}_{I_n(a)} =
\begin{cases} 
\triangle J_{I_n(a)}(1) & \text{if } \triangle J_{I_n(a)}(1) \geq C_{1_{n,a}} \\ 
\triangle J_{I_n(a)}(2) & \text{if }\triangle J_{I_n(a)}(2) \geq C_{2_{n,a}} \\ 
   ..          &    ..  \\
   \triangle J_{I_n(a)}(K_{n,a}) & \text{if }\triangle J_{I_n(a)}(K_{n,a}) \geq C_{K_{n,a}} \\ 
   0                 & \text{if none of above satisfies}
   
\end{cases}
\end{equation}
where the financial thresholds $\{C_{1_{n,a}},C_{2_{n,a}},..., C_{K_{n,a}} \}$ are set by the customers as described above. 

As a result, the optimal waiting time of interruptible appliance $a$ for customer $n$ can be obtained by solving the following optimization problem: 
\vspace{-0.4cm}
\begin{align}\label{waiting1}
\max_{k_{n,a}} \triangle \hat{J}_{I_n(a)}
\end{align}
Similarly, the waiting time cost model for non-interruptible appliances can be represented as the following optimization problem:
\vspace{-0.35cm}
\begin{align}\label{waiting2}
\max_{k_{n,a} } \triangle \hat{J}_{NI_n(a)}
\end{align}

\subsubsection{Curtailable Appliances}
Similarly to the interruptible and non-interruptible appliances, the customers can set the valid scheduling window $ \mathcal{H}_{n,a} \triangleq \{\alpha_{n,a}, \alpha_{n,a} + 1,...,\beta_{n,a}\}$.  However, compared with interruptible and non-interruptible appliances, the scheduling window of curtailable appliances should be more strict and accurate because the appliances will be \lq on\rq \text{ }for the whole window.


By the customers' behaviour analysis (i.e. some customers are price sensitive and some others are less price sensitive), there are two types of optimization models for curtailable appliances and a customer can choose one of them dependent on his/her preference.



\paragraph{Minimize Bill Subject to an Acceptable Energy Consumption}

This optimization scheme targets price sensitive customers. The proposed optimization model is given below:\vspace{-0.15cm}
\begin{align}
&{\operatorname{\min  }} J1_{C_n(a)}(\alpha_{n,a} \colon \beta_{n,a}) = \underset{x_{n,a}^h} {\operatorname{\min}} \sum _{h=\alpha_{n,a}}^{\beta{n,a}} p^h \times x_{n,a}^h \label{10} \\
&s.t. \nonumber \\
& \underline{u}_{n,a}^h   \leq x_{n,a}^h \leq \overline{u}_{n,a}^h , \label{11}\\
& \sum_{h=\alpha_{n,a}}^{\beta{n,a}} x_{n,a}^h \geq U_{n,a}^{min}. \label{12}
  \end{align}
Constraint (\ref{11}) shows that the energy consumption at each hour is within the minimum acceptable consumption level $\underline{u}_{n,a}^h$  and maximum affordable consumption level $\overline{u}_{n,a}^h$, which can be set according to each individual customer's preferences. Constraint (\ref{12}) indicates that the electricity consumed during the operation period should not be less than a minimum acceptable consumption level, i.e. there exists an energy consumption constraint for each curtailable appliance.

\paragraph{Maximize Energy Consumption Subject to an Affordable Financial Constraint}
This optimization scheme aims at the less price sensitive customers who prefer a budget based energy consumption maximization model. The proposed optimization model is given as follows:\vspace{-0.5cm}
\begin{align}
&{\operatorname{\max  }} J2_{C_n(a)}(\alpha_{n,a} \colon \beta_{n,a}) = \underset{x_{n,a}^h} {\operatorname{\max}} \sum _{h=\alpha_{n,a}}^{\beta{n,a}} x_{n,a}^h  \label{13} \\
&s.t. \nonumber\\
&\underline{u}_{n,a}^h   \leq x_{n,a}^h \leq \overline{u}_{n,a}^h , \label{14}\\
&\sum_{h=\alpha_{n,a}}^{\beta{n,a}} p^{h} \times x_{n,a}^h \leq C_{n,a}^{max}. \label{15}
  \end{align}
   
Constraint (\ref{14}) is same as constraint (\ref{11}). Constraint (\ref{15}) indicates that for each curtailable appliance, the money spent during the operation period should not exceed the given budget, i.e. there exists a financial cap for each curtailable appliance.
 



Since there are two types of optimization models for curtailable appliances, the optimization problem for customer $n$ including all types of appliances has two different optimization objectives shown as Eqs.(\ref{eq16}) and (\ref{eq17}). The customers can choose one of them depending on their preferences. 
\begin{equation} \label{eq16}
\begin{array}{l} 
{\operatorname{\min}} J1_n = {\operatorname{\min  }} \{ \underset{a \in I_n} \sum  (J_{I_n(a)} - \triangle \hat{J}_{I_n(a)}) + \\
\underset{a \in NI_n} \sum (J_{NI_n(a)} -  \triangle \hat{J}_{NI_n(a)}) + \underset{{a \in C_n}} \sum J1_{C_n(a)}\}\\

\text{subject to constraints \eqref{2}--\eqref{3}, \eqref{6}--\eqref{8}, and \eqref{11}--\eqref{12}}.
 \end{array}
\end{equation}
\begin{equation} \label{eq17}
\begin{array}{l} 
{\operatorname{\min}} J2_n = {\operatorname{\min  }} \{\underset{a \in I_n} \sum  (J_{I_n(a)} - \triangle \hat{J}_{I_n(a)} ) + \\
\underset{a \in NI_n} \sum (J_{NI_n(a)} -  \triangle \hat{J}_{NI_n(a)}) - \underset{{a \in C_n}} \sum J2_{C_n(a)}\}\\
 \text{subject to constraints \eqref{2}--\eqref{3}, \eqref{6}--\eqref{8}, and \eqref{14}--\eqref{15}}.
 \end{array}
\end{equation}
%

%
%
%

\subsection{Retailer-side Problem at the Upper Level}

In this subsection, we model the profit of the retailer by using the revenue subtracting the energy cost imposed on the retailer.

We define a cost function $ C_h(L_h)$ indicating the cost of the retailer providing electricity at each hour $h \in \mathcal{H} $, where $ L_h $  represents the amount of power provided to all customers at each hour of the day. We assume that the cost function $ C_h(L_h)$ is convex increasing in $ L_h $ for each $h$ \cite{Mohsenian-Rad2010} \cite{Li2011}. In view of this, the cost function is designed as follows \cite{Mohsenian-Rad2010}.
\begin{equation} \label{costfunction}
C_h(L_h) = a_h L_h^2 + b_h L_h + c_h 
\end{equation}
where $a_h > 0$ and $b_h \geq 0, c_h \geq 0$ at each hour $h \in \mathcal{H}$.

As a result, the profit maximization model is given as follows:
\begin{align}
&\underset{p^h} {\operatorname{\max}} \{ \sum\limits_{h \in \mathcal{H}} {{p^h}}\times\sum\limits_{n \in \mathcal{N}} {\sum\limits_{a \in {A_n}} {x_{n,a}^h} } - \sum\limits_{h \in \mathcal{H}} {{C_h}(\sum\limits_{n \in \mathcal{N}} {\sum\limits_{a \in {A_n}} {x_{n,a}^h} } )\} } \label{18} \\
&s.t. \nonumber \\
&p_h^{min} \leq p^h \leq p_h^{max}, \label{19}\\
&\sum\limits_{n \in \mathcal{N}} {\sum\limits_{a \in {A_n}} {x_{n,a}^h} }  \le E_h^{\max },\forall h \in \mathcal{H} , \label{20}\\
& \sum\limits_{h \in \mathcal{H}} {{p^h}}\times\sum\limits_{n \in \mathcal{N}} {\sum\limits_{a \in {A_n}} {x_{n,a}^h} } \leq R^{max} \label{21}. 
  \end{align}

Constraint (\ref{19}) represents that the prices the retailer can offer are greater than a minimum price, for example, the wholesale price at each hour, and less than a maximum price, for example, the price cap of the retail price due to retail market competition and regulation. Constraint (\ref{20}) indicates that there usually exists a maximum supply capacity by the retailer or a maximum load capacity of power networks. Due to the in-elasticity of energy use, we add the revenue constraint (\ref{21}) to improve the  acceptability of the retailer's pricing strategies, i.e. there exists a total revenue cap, denoted as $ R^{max}$, for the retailer. Without such a constraint, the retail prices will keep going up to a level which is politically against the government, political parties, and energy regulators and financially unacceptable by customers. 

%
%
%

\section{Bilevel Model Solutions} \label{model_solution}


In this paper, we propose a hybrid optimization approach to solve the bilevel optimization problem. Our approach determines the energy prices by interacting with the customers (smart meters) within the framework of a genetic algorithm for the upper level problem and individual optimization algorithm for the lower level problem. 

In this section we will firstly prove the existence of an optimal solution to our bi-level model, secondly show the solution algorithm to the lower-level problem, and finally present the solution algorithm to the upper-level problem.

\subsection{Existence of Optimal Solutions to the Bilevel Model} \label{existence}
First, we consider the following bilevel model with one upper level agent and one lower level agent. \\

\vspace{-1cm}
\begin{equation}  \label{bilevel_formualtion2}
\begin{array}{ll} 
\max \limits_{x,y_{1},...,y_N}& F(x,y_{1},...,y_N) \\ 
\hbox{subject to} & \\ 
& (y_1,...y_N) \in \underset{y_1,...y_N} {\operatorname{argmin}} \lbrace \hspace{1mm} \sum_{i=1}^N f_i(x,y_i): \\
& \quad \quad \quad \quad \quad \quad g_i(x,y_i)\leq 0, i=1,\dots,N \hspace{1mm} \rbrace\\ & G(x,y_{1},...,y_N)\leq 0 \\ & x\in X, (y_1,..., y_N) \in Y_1 \times ... \times Y_N 
\end{array}
\end{equation}

Note that each $f_i(x,y_i) \hspace{0.1cm} (i =1,...,N)$ in the objective function of the above lower level problem is independent from each other. Each constraint function $g_i(x,y_i) \hspace{0.1cm} (i =1,...,N)$ of the lower level problem is also independent from each other.

Further it is always assumed that the above considered bilevel optimization problem has at least one feasible solution.

\begin{lemm} \label{lemma} 
The bilevel model with one upper level agent and $N$ independent lower level agents (Eq.(\ref{bilevel_formualtion})) is equivalent to the bilevel model with one upper level agent and one lower level agent (Eq.(\ref{bilevel_formualtion2})).
\end{lemm}

The proof of Lemma \ref{lemma} is given in Appendix \ref{appendix A}. 


\begin{lemm} \label{lemma1}
Consider the bilevel model shown as Eq.(\ref{bilevel_formualtion2}), if $X$ is a finite space, then the optimal solutions to the bilevel model exist. 
\end{lemm}

The detailed proof of Lemma \ref{lemma1} is given in Appendix \ref{appendix B}. 

\begin{theo} \label{theoB1}
Consider the bilevel model with one upper-level decision agent (retailer) shown as  Eqs.(\ref{18} - \ref{21}) and $N$ independent lower-level decision agents (customers) shown as  Eq.(\ref{eq16}) or Eq.(\ref{eq17}). Then an optimal solution to the bilevel model exists. 
\end{theo}

The proof of Theorem \ref{theoB1} is given in Appendix \ref{appendix C}.

\subsection{Solutions to the Lower-level Problem} \label{4-a}

As the lower-level optimization problem is the sum of three separable sub-optimization problems corresponding to interruptible, non-interruptible, and curtailable appliances respectively, the lower-level problem can be solved by solving each sub-optimization problem separately.  

\subsubsection{ Interruptible Appliances}

The mathematical model of interruptible appliances is shown as Eqs.(\ref{1} - \ref{3}) which is a typical linear programming problem and can be solved using the optimization software. 

\subsubsection{Non-interruptible Appliances} 

We firstly define the sub-problem of the original model Eqs.(\ref{5} - \ref{8}) for non-interruptible appliances as follows by fixing the start time at $s'_{n,a} \in [\alpha_{n,a}, \beta_{n,a} - L_{n,a}]$.
\begin{equation} \label{basic-non-interruptible}
\begin{array}{l} 
   \underset{x_{n,a}^h} {\operatorname{\min}} \sum_{h=s'_{n,a}}^{s'_{n,a}+ L_{n,a}}  p^h \times  x_{n,a}^h\\
    s.t. \\
    
 \sum_{h=s'_{n,a}}^{s'_{n,a}+ L_{n,a}} x_{n,a}^h=E_{n,a},\\
 
 \gamma_{n,a}^{min}\leq x_{n,a}^h \leq \gamma_{n,a}^{max}, \forall h \in [s'_{n,a}, s'_{n,a}+ L_{n,a}].
 \end{array}
 \end{equation}

Eq.(\ref{basic-non-interruptible}) is a linear programming problem and can be solved by the optimization software. As a result, the original problem Eqs.(\ref{5} - \ref{8}) can be solved in an iterative manner. 

\subsubsection{Curtailable Appliances}

The optimization problems Eqs.(\ref{10} -\ref{12}) and Eqs.(\ref{13} - \ref{15}) for curtailable appliances are linear programming problems and can be solved by the optimization software.

\subsection{Distributed Algorithms to the Upper Level Problem}

Due to the existence of the starting time and waiting time in the lower-level problem, which makes the lower-level problem non-differential and  discontinuous, in this subsection, we adopt GA based distributed optimization algorithms to solve the profit maximization problem at the retailer's side and show how the retailer finds the optimal electricity day-ahead prices by taking into account the customers' responses. 



To avoid too much data passing between the retailer and the smart meters and reduce the number of generations for the GA (each generation needs to pass a new group of prices distributedly to all customers/agents to re-compute their optimal reactions which is very costly), we propose two strategies that improve the algorithms' performance : 1) use a larger population for the GA. This strategy is based on the observations that the local optimization problems (customers side) are simple and easy to compute, even with a very large population; 2) reduce corresponding GA generations to improve the algorithm efficiency as such a large population size can ensure the GA's convergence. 

\begin{algorithm}[!t]

\renewcommand{\algorithmicrequire}{\textbf{}}

\caption{Multi-population GA based pricing algorithm to Eqs.(\ref{18} - \ref{21}) executed by the retailer}
\label{Algorithm:GA based decision-making scheme algorithm}

\begin{algorithmic}[1]

\STATE Population Initialization, i.e. generating a population of $N$ chromosomes randomly. 

\STATE Produce $C$ sub-populations, i.e. each sub-population has $N/C$ individuals.
\STATE Each sub-population evolves in a traditional GA way shown as steps (4 -- 9).
\FOR{ i=1 to $N/C$  } 

\STATE The utility company announces strategy $i$, i.e. it announces a set of 24-hour prices by decoding the $i$th chromosome to the smart meters (customers) via two way communication infrastructure.

\STATE Receive the optimal response of each customer $n$ (smart meter) including the optimal energy consumption information to strategy $i$.

\STATE Check the feasibility of strategy $i$ to see if it satisfies all the constraints Eqs.(\ref{19} - \ref{21}). If not, handle the invalid individuals by the approach proposed in \cite{deb2000efficient}. Then, obtain the fitness value of  strategy $i$. 
\ENDFOR

\STATE A new generation of chromosomes is created by using the selection, crossover and mutation operations.

\STATE Migrations between sub-populations.

\STATE Steps (3 - 10) are repeated until the stopping condition is
reached.

\STATE The retailer announces the final price vector to the smart meters (customers) via LAN at the beginning of the scheduling horizon.
\end{algorithmic}
\end{algorithm}


\begin{algorithm}[!t]

\renewcommand{\algorithmicrequire}{\textbf{}}

\caption{Energy management system executed by each smart meter (customer)}
\label{Energy scheduling algorithm}

\begin{algorithmic}[1]

\STATE Receive the price information from the retailer.

\STATE The smart meter calculates the energy consumption in response to prices by solving the lower-level problem Eq.(\ref{eq16}) or Eq.(\ref{eq17}).

\STATE The smart meter sends back the total energy consumption at each hour to the retailer via two way communication infrastructure. 

\end{algorithmic}
\end{algorithm}

Instead of simply increasing the population size, we propose a multi-population GA method \cite{muhlenbein1991parallel} to tackle the problem, i.e. a single population is divided into multiple sub-populations and each sub-population evolves in a traditional GA way. In addition, the individuals migrate from one sub-population to another from time to time, known as the island model \cite{whitley1999island} and we use the ring migration type topology where individuals are 
transferred between directionally adjacent sub-populations \cite{tang1996genetic}. In the GA setting for each sub-population, binary encoding and deterministic tournament selection without replacement is adopted. For the crossover and mutation operations, we employ uniform crossover and bit flip mutation respectively. The constraints for the upper level problem are handled by the approach proposed in \cite{deb2000efficient}.

%

%

Finally, the multi-population GA based distributed algorithms are shown in Algorithm \ref{Algorithm:GA based decision-making scheme algorithm} and \ref{Energy scheduling algorithm}, which are implemented at the retailer-side and customer-side respectively. 


At the end, the most profitable prices for the retailer and the best usage patterns and schedules with the maximized benefits for each customer are found.

\section{Numerical Results} \label{results}

We simulate a neighbourhood consisting of 100 customers served by one energy retailer. It is assumed that each customer has 4 appliances: PHEV, dishwasher, washing machine and air-conditioning. The scheduling horizon is set from 8AM to 8AM (the next day). We consider heterogeneous customers, i.e. the customers are different in terms of energy use and appliance settings. In the following, we give the parameter settings for both the lower-level model and the upper-level model.

Note that, in the following, $\alpha_{n,a}$ and $\beta_{n,a}$ are uniformly distributed integers for all appliances settings.


For PHEV, recall from Eqs.(\ref{1} -- \ref{3}), $E_{n,a}$ is chosen from the uniform distribution on [9, 11] kWh. Each $\gamma_{n,a}^{min}$ is $0$ kWh, and each $\gamma_{n,a}^{max}$ is chosen  from the uniform distribution on [2.5, 3.3] kWh. $\alpha_{n,a}$ is chosen from the uniform distribution on [6, 9] PM, and $\beta_{n,a}$ is chosen from the uniform distribution on [5, 8] AM (the next day).

For dishwasher, $E_{n,a}$ is chosen from the uniform distribution on [2.3, 2.9] kWh. Each $\gamma_{n,a}^{min}$ is $0$ kWh, and each $\gamma_{n,a}^{max}$ is chosen from the uniform distribution on [1.2, 1.7] kWh. $\alpha_{n,a}$ is chosen from the uniform distribution on [8, 11] AM, and $\beta_{n,a}$ is chosen from the uniform distribution on [6, 9] PM. 

For washing machine, $E_{n,a}$ is chosen from the uniform distribution on [1.8, 2.3] kWh. Each $\gamma_{n,a}^{min}$ is $0$ kWh, and each $\gamma_{n,a}^{max}$ is chosen from the uniform distribution on [1.0, 1.5] kWh. $\alpha_{n,a}$ is chosen from the uniform distribution on [6, 9] PM, and $\beta_{n,a}$ is chosen from the uniform distribution on [5, 8] AM (the next day).  

For the air-conditioning, for the purpose of simulations, we assume that all the customers choose the second optimization model, i.e. maximize energy consumption subject to an acceptable financial constraint. As a result, $\underline{u}_{n,a}^h$ is chosen from the uniform distribution on [0.5, 0.8] kWh, and $\overline{u}_{n,a}^h$ is chosen from the uniform distribution on [1.8, 2.2] kWh. $\alpha_{n,a}$ is chosen from the uniform distribution on [4, 6] PM, and $\beta_{n,a}$ is chosen from the uniform distribution on [9, 11] PM (the next day).
$C_{n,a}^{max}$ is chosen from the uniform distribution on [70, 90] pence. 

Furthermore, the upper bound of hourly energy consumption for each household $E_n^{max}$ is chosen from the uniform distribution on [3.5, 4.5] kWh. 




For the cost of the energy provided to customers by the utility company, we model this as a cost function shown as Eq.(\ref{costfunction}). We assume that $b_h = 0, c_h=0$ for all  $h\in \mathcal{H}$. Also, we have $a_h = 5.5 \times 10^{-4} $ pence during the day, i.e. from 8AM to 12AM and $a_h = 4.0 \times 10^{-4}$ pence at night hours, i.e. from 12AM to 8AM (the next day).

In this section, firstly the convergence analysis of our proposed algorithms is given. Secondly, we show the benefits to the retailer by employing our proposed day-ahead pricing scheme, which is compared with a flat pricing scheme. Thirdly, we present the benefits to customers by adopting our proposed energy management scheme.

\subsection{Convergence Analysis} \label{convergence analysis}

In this subsection, we test the aforementioned two strategies (i.e. increasing the population and reducing the generation) in terms of convergence for the GA. 

\begin{figure}[!t]
\centering
  \includegraphics[width=8cm]{{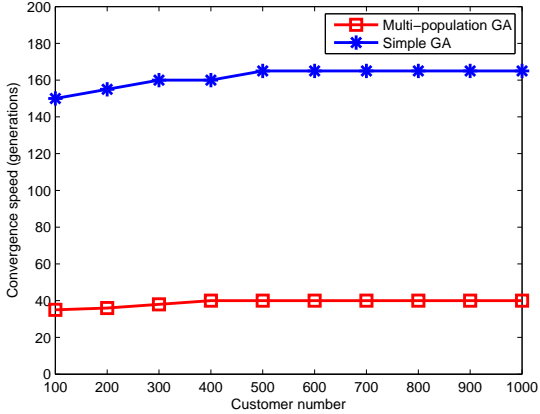}}
  \captionof{figure}{ Convergence of the multi-population GA and the simple GA under different customer numbers}
\label{user_different}
  \end{figure}

We conduct simulations to show the convergence speed of our proposed multi-population GA and the simple GA under different customer numbers (from 100 to 1000) as shown in Figure \ref{user_different}. It is worth mentioning that the convergence speed does not change much when customer numbers increase, which indicates our proposed distributed optimization algorithm is rather scalable with the number of customers and  can significantly reduce the data communication between the retailer and the smart meters.

\subsection{Benefits to the Retailer}

In this subsection, we compare our proposed optimal day-ahead pricing scheme with optimal flat pricing scheme. The parameter settings of our proposed multi-population GA are shown in Table \ref{Table:Parameter settings of mGA}.

Since the customers have no incentives to change their energy consumption pattern when responding to flat pricing, we assume that, under the flat pricing, the customers start the operations of appliances right at the beginning of the scheduling window $H_a$ and the appliances work at their typical power levels. We assume that, for each hour $h$, $ 8.0$ pence $\leq p^h \leq 14.0 $ pence holds.  When calculating the optimal flat pricing, we use the same parameters and model as those of optimal day-ahead pricing.

\begin{figure}[!t]
\centering
  \includegraphics[width=8cm]{{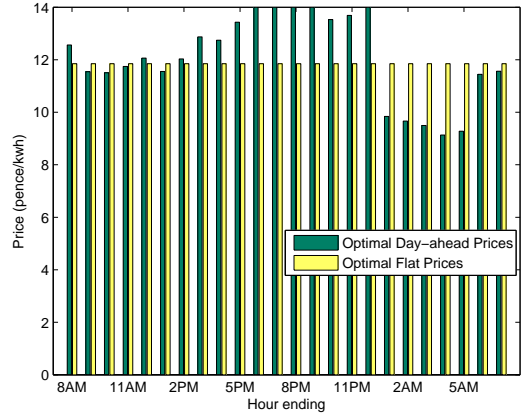}}
  \captionof{figure}{Obtained optimal day-ahead prices and flat prices}
  \label{rtp}
  \end{figure}

The obtained optimal day-ahead prices and flat prices are given in Figure \ref{rtp}. Finally, the details of  revenue, cost and profit under optimal day-ahead prices and flat prices can be found in Table \ref{Table:4}.

 \begin{table}[!t]
  \caption{Parameter settings of the multi-population GA }
  \label{Table:Parameter settings of mGA}  
  \centering 
  \begin{tabular}{p{3.5cm}p{2cm}p{2cm}}
       \hline\noalign{\smallskip}
      Parameter Name & Symbol  & Values\\
      \noalign{\smallskip}\hline\noalign{\smallskip}
      Number of Sub-population & $S_p$ & 15 \\
      Sub-population Size & $N$ & 40 \\
      Migration Rate & $M_r$ & 0.2 \\
      Chromosome Length & $L$ & 10\\
      Mutation Probability   & $P_m$ & 0.01 \\
      Terminate Generation & $ T$ &  100 \\
      \noalign{\smallskip}\hline
    \end{tabular}
\end{table}


From Table \ref{Table:4}, we can see that, to make the same revenue (i.e. the total bills for all customers are the same), the cost of the retailer under optimal day-ahead pricing is 120.08 pounds and the cost under optimal flat pricing is higher (139.35  pounds). This is due to the increase of peak demand and thus the increase of peak-time cost. Furthermore, the profit under optimal day-ahead pricing (134.92 pounds) is higher than the profit under optimal flat pricing (115.65 pounds). The example shows a very important potential for the day-ahead pricing and our proposed approach: the day-ahead pricing enables to increase the retailer's profit without increasing customers' expenses.
   

   \begin{table}[!t]
   \centering
  \caption{Revenue, cost and profit under different price}
     \label{Table:4} 
    \begin{tabular}{{p{3.0cm}p{1.2cm}p{1.2cm}p{1.2cm} }}
    \hline
      Price setting & Revenue (pounds) & Cost (pounds) & Profit (pounds) \\
\hline\noalign{\smallskip}      
      Optimal Day-ahead Pricing  & 255.00  & 120.08
 & 134.92	 \\
 
 Optimal Flat Pricing & 255.00  & 139.35  & 115.65 \\
\hline
\end{tabular}  
\end{table}

\subsection{Benefits to Customers}

In this subsection, we show the effectiveness of the proposed energy management scheme based on public day-ahead price data. We use the actual electricity prices data adopted by ISO New England from January 1, 2012 to January 31, 2012, which is available to the public on-line at \cite{newengland2013}. Due to the space limitation, we only show the result of the first customer. 


%

\begin{figure}[!t]
\centering
  \includegraphics[width=8cm]{{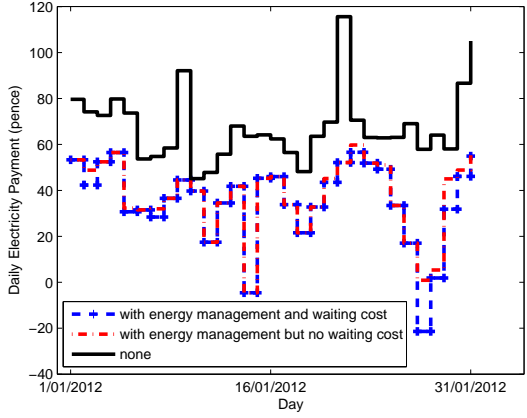}}
  \captionof{figure}{Daily electricity payment of one customer over one month}
  \label{bills}
  \end{figure}

The simulation result is shown as Figure \ref{bills} where we can find that, after adopting the energy management scheme, the daily bill payment is significantly reduced. Furthermore, we show that by adopting our proposed financial incentive based waiting time scheme, the customers may get further benefits in  terms of reducing their payments subject to acceptable life comforts.  

\section{Conclusion} \label{conclusion}
In this paper, we model the interactions between the retailer and its customers as a bilevel optimization problem. Firstly, according to the load types, we categorize home appliances into interruptible, non-interruptible and curtailable appliances. For different categories of appliances, different appliance-level optimization models are given, which forms the lower level problem. As the common solutions to the bilevel optimization problem such as KKT transformations are not usable in our application setting, a hybrid optimization approach based on genetic algorithms and individual optimization solutions  has been proposed to solve  the bilevel problem. Since the numerical results show that both the retailer and its customers can benefit from the proposed model, it has great potential to improve the implementations of current energy pricing programs, help customers to reduce their increasing energy bills, and change their energy usage patterns.
%

\appendices

\section{Proof of Lemma \ref{lemma}} \label{appendix A}
\begin{proof}
For any fixed $x$, as 
\begin{align*}
\underset{(y_1,...,y_N) \in Y_1\times ...\times Y_N} {\operatorname{\min}} \lbrace \hspace{1mm} \sum_{i=1}^N f_i(x,y_i): g_i(x,y_i)\leq 0, \hspace{0.2cm} \\ i=  1,...,N \rbrace  = \sum_{i=1}^N \underset{y_i \in Y_i}{\operatorname{\min}} \lbrace \hspace{1mm} f_i(x,y_i):g_i(x,y_i)\leq 0\rbrace
\end{align*} 
which implies immediately that 
\begin{align*}
(y_1^*,...,y_N^*) \in \underset{(y_1,...,y_N) \in Y_1\times ...\times Y_N} {\operatorname{argmin}} \lbrace \hspace{1mm} \sum_{i=1}^N f_i(x,y_i): \\ g_i(x,y_i)\leq 0,  i=1,\dots,N \hspace{1mm} \rbrace
\end{align*} 
 if and only if \begin{align*}
  y_i^* \in \underset{y_i \in Y_i}{\operatorname{argmin}} \lbrace \hspace{1mm} f_i(x,y_i): g_i(x,y_i)\leq 0\rbrace \quad i=1,2...,N.
  \end{align*} 
Based on the formulations of Eq.(\ref{bilevel_formualtion}) and Eq.(\ref{bilevel_formualtion2}), this implies  that they have the exact same objective functions and constraints and therefore these two bilevel optimization problems are equivalent and have the same optimal solutions. 
\end{proof}

\section{Proof of Lemma \ref{lemma1}} \label{appendix B}

\begin{proof}
Firstly, for a given $x \in X$, denote  
\begin{equation} \label{lowerlevel}
\begin{array}{cc}
\Omega(x) = \underset{y_1,...,y_N \in Y_1\times ...\times Y_N} {\operatorname{argmin}} \lbrace \hspace{1mm} \sum_{i=1}^N f_i(x,y_i): \\ \quad \quad \quad \quad g_i(x,y_i)\leq 0,  i=1,\dots,N \rbrace,
\end{array}
\end{equation}
and choose 
\begin{equation} \label{upperlevel1}
\begin{array}{cc}
(y_1^*,...,y_N^*) = \underset{y_1,...,y_N \in \Omega(x)} {\operatorname{argmax}} \{ F(x,y_1,...,y_N) : \\ \quad \quad \quad \quad \quad \quad \quad \quad \quad \quad   G(x,y_1,...,y_N) \leq 0 \}.
\end{array}
\end{equation}

Now denote $(y_1^*,...,y_N^*) = R(x)$. As $X$ is a finite set, there exists an optimal solution as follows
\begin{equation} \label{upperlevel2}
\begin{array}{cc}
x^* = \underset{x \in X} {\operatorname{argmax}} \left\lbrace F[x,R(x)] : G[x,R(x)] \leq 0 \right\rbrace.
\end{array}
\end{equation}

Now we are going to prove $[x^*, R(x^*)]$ is the optimal solution to the bilevel problem \eqref{bilevel_formualtion2}.

For any feasible solution $(x,y_1,...,y_N)$ to the bilevel problem \eqref{bilevel_formualtion2}, from \eqref{lowerlevel}, \eqref{upperlevel1}, \eqref{upperlevel2}, we have
\begin{equation} \label{upperlevel3}
\begin{array}{cc}
F(x,y_1,...,y_N)  \leq F[x,R(x)] \leq F[x^*,R(x^*)].
\end{array}
\end{equation}\
As $[x^*, R(x^*)]$ is a feasible solution to \eqref{bilevel_formualtion2} based on \eqref{lowerlevel}, \eqref{upperlevel1}, \eqref{upperlevel2}, the inequality \eqref{upperlevel3} implies that the objective function of the bilevel problem \eqref{bilevel_formualtion2} takes its maximal value at $[x^*, R(x^*)]$. Therefore, $[x^*, R(x^*)]$ is the optimal solution to \eqref{bilevel_formualtion2}.
\end{proof}

\section{Proof of Theorem \ref{theoB1}} \label{appendix C}

\begin{proof}
Firstly, according to Lemma \ref{lemma}, our considered bilevel pricing optimization problem given in \eqref{eq16}, \eqref{eq17} and \eqref{18}-\eqref{21} with one upper-level decision agent (retailer) and $N$ independent lower-level decision agents (customers) is equivalent to the bilevel model with one upper-level decision agent and one lower-level decision agent. Therefore we only need to prove the existence of the optimal solution in the formation of the optimization problem as \eqref{bilevel_formualtion2}. 


For each decision variable $p^{h} \; (h=1,...,24)$ in decision variable vector $P =\left(p^{1} ,...,p^{24} \right)$ at the upper level problem, it only takes finite values (in practice it is often one decimal after the small unit in a given currency), i.e., the price at each hour $h$ can only take $D_{h}$ values, where $D_{h}$ is the number of possible price values within interval $[p_{h}^{\min }, p_{h}^{\max }]$. As a result, $D=\prod_{h=1}^{24}D_{h}$ is a finite integer. 

We denote the space of all prices across 24 hours (i.e., $P$) at the upper-level problem as $U_P$. Noting the total number of elements of set $U_P$ is $D$, which is a finite integer, it is implied immediately that $U_P$ is a finite space. Therefore, based on Lemma \ref{lemma1}, it is implied that the optimal solution to the considered bilevel pricing problem exists.

\end{proof}

\section*{Acknowledgment}

This work is partly supported by the National Nature Science Foundation of China (Grant No. 71301133) and Humanity and Social Science Youth Foundation of Ministry of Education, China (Grant No. 13YJC630033).

\balance



\bibliographystyle{IEEEtran}
\bibliography{Smart}

\end{document}